\documentclass[authoryear,round,12pt]{article}
\usepackage{color}
\usepackage{pslatex}
\usepackage{graphicx}
\usepackage{setspace}
\usepackage{natbib}
\usepackage{amsmath}
\usepackage{subfigure}
\newcommand{\beq}{\begin{equation}}
\newcommand{\eeq}{\end{equation}}
\newcommand{\myabstract}{This document is part of a series of near
  real-time weekly influenza forecasts made during the 2012-2013
  influenza season.  Here we present results of a forecast initiated
  following assimilation of observations for Week 51 (i.e. the
  forecast begins December 23, 2012) for municipalities in the United
  States.  The forecast was made on December 28, 2012.  Results from
  forecasts initiated the four previous weeks (Weeks 47-50) are also
  presented.  Predictions generated with an alternate SIRS model, run
  without absolute humidity forcing (no AH), are also presented.}

\newcommand{\myacknow}{Funding was provided by US NIH grant GM100467
  (JS, AK, ML), as well as NIEHS Center grant ES009089 (JS) and
  the RAPIDD program of the Science and Technology Directorate, US
  Department of Homeland Security (JS).  The content is solely the
  responsibility of the authors and does not necessarily represent the
  official views of the National Institute Of General Medical
  Sciences, National Institutes of Health, or Department of Homeland
  Security.}

\begin{document}
%
%
\title{\textbf{\large{Week 51 Influenza Forecast for the 2012-2013
      U.S. Season}}}
%
%
\author{\textsc{Jeffrey Shaman}
                                \thanks{\textit{Corresponding author address:} 
                                Jeffrey Shaman, Department of
                                Environmental Health Sciences, Mailman
                                School of Public Health, Columbia
                                University, 722 West 168th Street,
                                Rosenfield Building, Room 1104C, New
                                York, NY 10032. 
                                \newline{E-mail:
                                  jls106@columbia.edu}}\quad\textsc{}\\
\centerline{\textit{\footnotesize{Department of Environmental Health Sciences,
    Mailman School of Public Health, Columbia University, New York, New York}}}
\and
\centerline{\textsc{Alicia Karspeck}} \\
\centerline{\textit{\footnotesize{Climate and Global Dynamics
      Division, National Center for Atmospheric Research, Boulder, Colorado}}}
\and 
\centerline{\textsc{Marc Lipsitch}} \\
\centerline{\textit{\footnotesize{Center for Communicable Disease
      Dynamics, Harvard School of Public Health, Harvard University,
      Boston, Massachussetts}}}
}

\maketitle

{
\begin{abstract}
\myabstract
\end{abstract}
}

\section{Background}
\label{sec:retrofore}

This record is part of an evolving series of real-time forecasts
developed during the 2012-2013 influenza season for the United States.
Additional documentation of earlier forecasts for this season have
also been posted \citep{Shaman-Karspeck-Lipsitch-2012:week49,
  Shaman-Karspeck-Lipsitch-2012:week50}.

Forecast skill is calculated for individual cities, as well as for
census divisions (regions) and all cities in aggregate, from
retrospective forecasts made for the 2003-2004 through 2011-2012
seasons, excluding the pandemic years 2008-2009 and 2009-2010 (which
will need to be handled separately in the future).  The forecast
methods are similar to those described in
\cite{Shaman-Karspeck-2012:forecasting}.  Based on the relationship
between prediction accuracy and ensemble spread of these retrospective
forecasts we can assign calibrated confidences to our current
predictions.

\begin{figure}[tbh]
\noindent\includegraphics[width=18pc,angle=0]{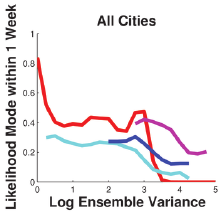}
\noindent\includegraphics[width=18pc,angle=0]{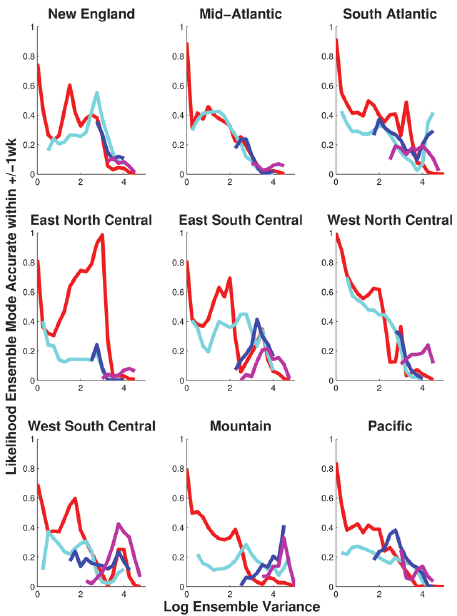}\\
\caption{Plot of ensemble mode forecast accuracy versus ensemble
  spread measured as log ensemble variance +1.  Left) 114 cities in
  aggregate.  The runs are binned in
  increments of 0.25 units and stratified by forecast lead time: 1-3
  weeks (red), 4-6 weeks (cyan), 7-9 weeks (blue), 10+ weeks
  (magenta).  Right) Same as left, but the 114 cities grouped by census region.}
\label{fig:allcities_mode_allclim}
\end{figure}

Figure \ref{fig:allcities_mode_allclim} shows this relationship
between prediction accuracy and ensemble spread for all cities in
aggregate using an SIRS model with climatological AH forcing and a
factor of 5 mapping.  Overall the relationship is informative;
however, for all lead times there is a basic plateau of skill once the
ensemble log variance drops below 2.5 to 3 weeks$^2$.

When the cities are grouped by region, there is some heterogeneity.
Some regions (e.g. West North Central) show marked improvement of
forecast accuracy/skill with decreasing spread across all lead times.
Other regions show much more limited skill--the Mountain region only
has skill at 1-3 weeks, and the East North Central has problems at 1-3
weeks.  
 
\section{2012-2013 Forecast}
\label{sec:actualfore}

\begin{figure}[tbh]
\noindent\includegraphics[width=20pc,angle=0]{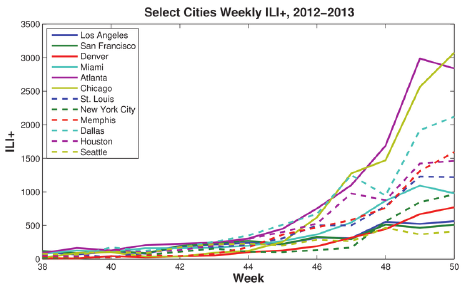}
\noindent\includegraphics[width=20pc,angle=0]{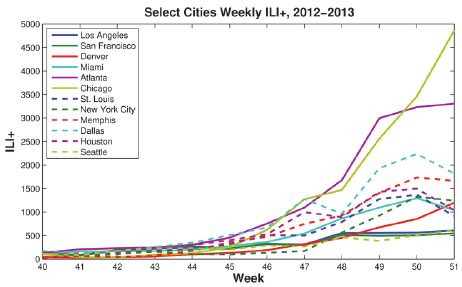}\\
\caption{Time series of: Left) Week 50 estimates of Weeks 38-50 ILI+
  and Right) Week 51 estimates of Weeks 40-51 ILI+ for the 2012-2013
  season.  ILI+ is Google Flu Trends weekly municipal ILI estimates
  times CDC census division influenza positive rates.}
\label{fig:selcitytimeseries}
\end{figure}

The forecasts are in fact complicated as the target--i.e. the ILI+
observations used for assimilation--is changing.  That is, the CDC
census division influenza positive tests are preliminary, and can (and
do) change as more reports are received.  As a result of this, the
observed peak may actually change.  This shifting can be seen in
comparison of the ILI+ plotted last week (with information through
Week 50) versus the current week (with information through Week 51,
Figure \ref{fig:selcitytimeseries}).  For instance, last week for
Atlanta, ILI+ during Week 49 was clearly higher than during Week 50.
This week, the numbers have been revised and Week 50 is higher than
Week 49 (and Week 51, is ever so slightly higher than Week 50).

\subsection{Week 51 Forecast}
\label{subsec:actual51}

\begin{figure}[tbh]
\noindent\includegraphics[width=25pc,angle=0]{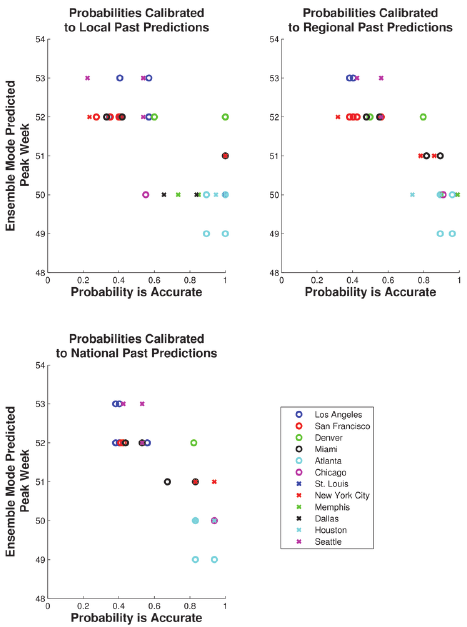}\\
\caption{Ensemble mode peak week predictions initiated December 23,
  2012, following assimilation of Week 51 observations, for 12 cities
  plotted as a function of probability/confidence calibrated from
  historical city, regional and national prediction accuracy.}
\label{fig:select_wk51fore_cal} 
\end{figure}

\begin{figure}[tbh]
\noindent\includegraphics[width=18pc,angle=0]{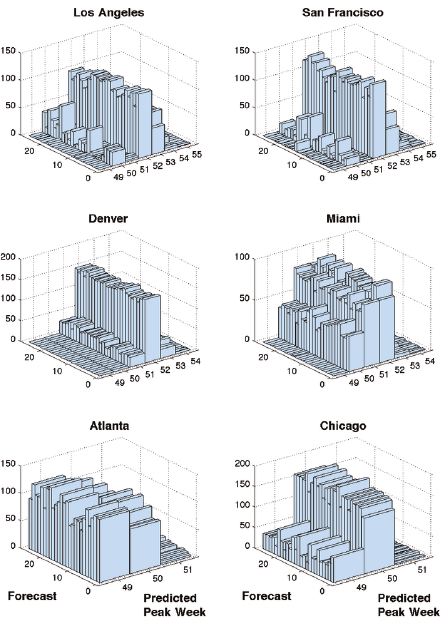}
\noindent\includegraphics[width=18pc,angle=0]{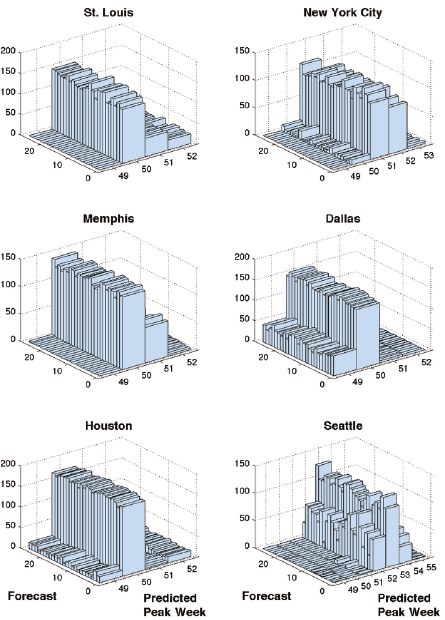}\\
\caption{Histograms of peak predictions from 25 200-member ensemble
  forecasts made beginning the start of Week 52 (December 23, 2012)
  for select cities.  The distributions show the ensemble spread among
  peak predictions.}
\label{fig:select_wk51fore}
\end{figure}

\begin{figure}[tbh]
\noindent\includegraphics[width=25pc,angle=0]{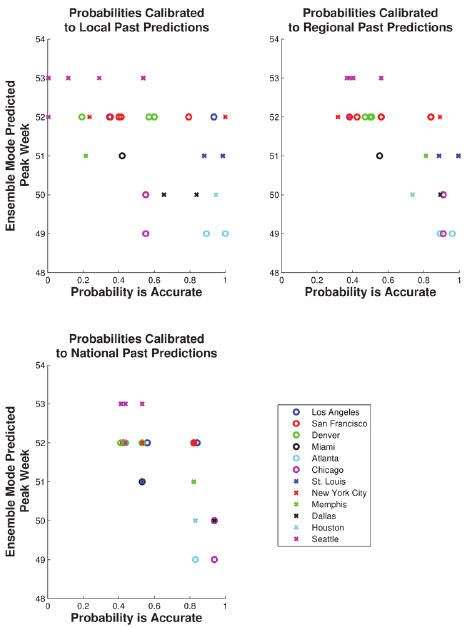}\\
\caption{Ensemble mode peak week predictions initiated December 16,
  2012, following assimilation of Week 50 observations, for 12 cities
  plotted as a function of probability/confidence calibrated from
  historical city, regional and national prediction accuracy.}
\label{fig:select_wk50fore_cal} 
\end{figure}

The Week 51 forecast (initiated December 23, 2012) predicts peaks for
Atlanta Week 49-50, 1-2 week in the past (Figure
\ref{fig:select_wk51fore_cal} and \ref{fig:select_wk51fore}).  This is
shifted from the prior week forecast (Week 50), which was firmly on
Week 49 (Figure \ref{fig:select_wk50fore_cal}).  Of course at that
time, Week 49 was higher than Week 50, and now it is the reverse.

\begin{figure}[tbh]
\noindent\includegraphics[width=18pc,angle=0]{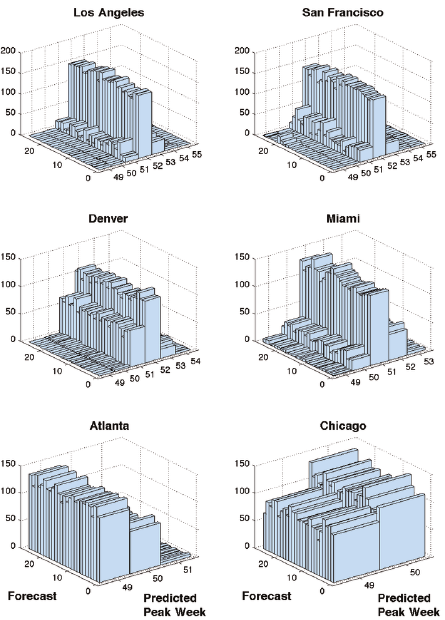}
\noindent\includegraphics[width=18pc,angle=0]{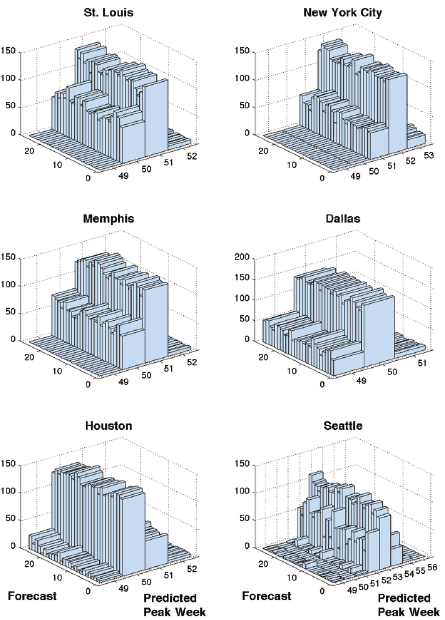}\\
\caption{Histograms of peak predictions from 25 200-member ensemble
  forecasts made beginning the start of Week 51 (December 16, 2012)
  for select cities.  The distributions show the ensemble spread among
  peak predictions.}
\label{fig:select_wk50fore}
\end{figure}

Basically, the distribution for Atlanta has shifted slightly from
consistently favoring Week 49 (Figure \ref{fig:select_wk50fore}) to a
more even split between predictions of peak during Week 49 and Week 50
(Figure \ref{fig:select_wk51fore}).  However, as Week 51 is currently
higher for ILI+ than either Week 49 and 50, unless the peak shifts
back, the Atlanta forecasts of Week 49 (including prior weeks) will be
off more than the $\pm1$ margin of error.

The current forecast for Chicago is for a peak during Week 50 (Figure
\ref{fig:select_wk51fore_cal}).  This is a slight shift from the Week
49-50 peak prediction made last week (Figure
\ref{fig:select_wk50fore_cal}).  Given that the Week 51 ILI+
observation is much higher than Week 50, predictions of peak during
Week 49 are outside the margin of error.  Again, the forecasts are
further challenged by the changes in prior week observations;
Chicago's Week 50 ILI+ changed quite a bit from last week to this,
having increased more than 12\%, which may affect model optimization
and prediction of the evolving trajectory of the outbreak.

Dallas is again predicted to peak during Week 50, as
are Memphis, St. Louis, and Houston.  These are all in line with the
prior weeks predictions.  All 4 of these cities \textit{currently}
peak during Week 50.

Both New York City and Miami have predicted peaks during Weeks 51 and
52 (the week past or the current week ending December 29, 2012).
These are roughly similar to the prior weeks predictions (Table
\ref{table:t1}).  In both cases, these are slight shifts in the
distribution of predictions from the prior week.  And in both cases,
the Week 51 prediction has greater confidence.  Miami currently
appears to peak during Week 50, New York is still rising (Figure
\ref{fig:selcitytimeseries}).

\begin{table}[t]
\caption{Summary of weekly model predictions at 12 select cities.  Weeks
   are labeled consecutively (Week 1 of 2013 is Week 53, etc.).
   Predictions were initiated at the end of Weeks 47, 48, 49, 50 and 51.
   The range of prediction confidences, derived from municipal,
   regional and national calibrations, are given in parentheses.}\label{table:t1}
 \begin{center}
\small
 \begin{tabular}{ccccrrcrc}
   \hline\hline
   City & Week 51 & Week 50 & Week 49 & Week 48 & Week 47\\
   & Prediction & Prediction & Prediction & Prediction & Prediction \\
   \hline
   Los Angeles &51-52 (35-60\%)&52 (50-95\%) & 51-52 (35-90\%) & 51-52 (20-55\%) & 51 (15-30\%) \\
   San Francisco &52 (25-45\%)& 52 (35-85\%) & 51-52 (25-40\%) & 51 (30-85\%) &
   50-51 (25-60\%)  \\
   Denver &52 (50-99\%)& 52 (20-60\%) & 52 (20-55\%) & 51-52 (0-55\%) & 51 (10-30\%) \\
   Miami &51-52 (30-99\%)& 51 (40-60\%) & 51 (40-99\%) & 50-51 (40-55\%) & 50-51 (0-45\%) \\
   Atlanta &49-50 (80-99\%)& 49 (80-99\%) & 49 (90-99\%) & 49 (80-95\%) & 49 (80-95\%) \\
   Chicago &50 (55-95\%)&49-50 (55-95\%) & 49 (55-95\%) & 49 (35-80\%) & 49 (35-80\%) \\
   St. Louis &50 (80-99\%)& 51 (85-99\%)  & 50-51 (80-99\%) & 50 (85-99\%) & 51 (30-90\%) \\
   New York City&51-52 (20-99\%) &52 (25-99\%) & 51 (25-99\%) & 52-53 (25-60\%) & 53-54
   (25-55\%)\\
   Memphis &50 (70-99\%)&51 (20-80\%) & 50 (20-80\%) & 50 (15-80\%) & 49-50 (15-55\%) \\
   Dallas &50 (65-95\%) &50 (65-90\%) & 49-50 (65-85\%) & 49 (50-75\%) & 49 (40-85\%) \\
   Houston &50 (70-95\%)&50 (75-90\%) & 50 (50-60\%) & 50 (50-60\%) & 49 (50-85\%) \\
   Seattle &51-52 (20-60\%)&52-53 (0-55\%) & 52-53 (5-55\%) & 51-52 (5-55\%) & 51 (5-35\%) \\
   \hline
 \end{tabular}
 \end{center}
\end{table}
\normalsize

Denver and San Francisco are now predicted to peak during Week 52.
These are consistent with the prior week forecasts (Table
\ref{table:t1}).  Los Angeles and Seattle are predicted to peak during
Weeks 52 and 53.  This is the same as before for Seattle, but a drift
farther into the future for Los Angeles.

\subsection{Week 51 Forecast -- No AH}
\label{subsec:actual51noAH}

\begin{figure}[tbh]
\noindent\includegraphics[width=25pc,angle=0]{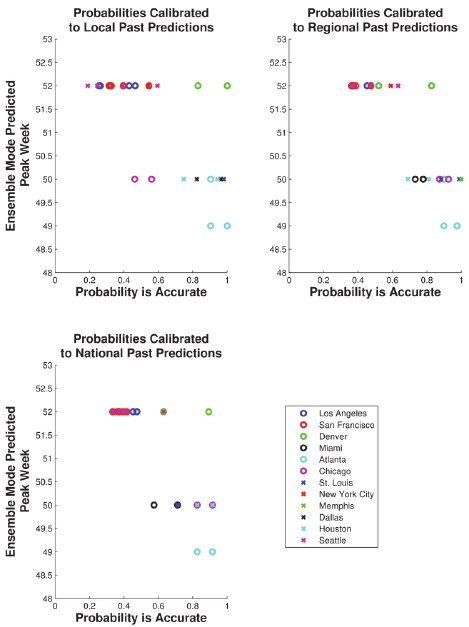}\\
\caption{Ensemble mode peak week predictions initiated December 23,
  2012, following assimilation of Week 51 observations using an SIRS
  model without AH forcing, for 12 cities plotted as a function of
  probability/confidence calibrated from historical city, regional and
  national prediction accuracy.}
\label{fig:select_wk51fore_noAH_cal}
\end{figure}

Forecasts initiated after assimilation of Week 51 observations at the
beginning of Week 52 (December 23, 2012) using an SIRS model without
absolute humidity forcing produce generally similar predictions
(Figure \ref{fig:select_wk51fore_noAH_cal}) as seen with the SIRS
model with climatological AH forcing.  The predictions for the 12
focus cities using the SIRS model without AH forcing are summarized in
Table \ref{table:t2}).

 \begin{table}[t]
   \caption{Summary of weekly model predictions at 12 select cities
     using an SIRS model without absolute humidity forcing.  Weeks
     are labeled consecutively (Week 1 of 2013 is Week 53, etc.).
     Predictions were initiated at the end of Weeks 48, 49, 50 and 51.
     The range of prediction confidences, derived from municipal,
     regional and national calibrations, are given in parentheses.}\label{table:t2}
 \begin{center}
 \begin{tabular}{ccccrrcrc}
 \hline\hline
 City & Week 51 &Week 50 & Week 49 & Week 48 \\
    & Prediction &Prediction & Prediction & Prediction \\
 \hline
  Los Angeles &52 (25-50\%)&51 (25-50\%) & 50-51 (25-50\%) & 50 (25-50\%) \\
  San Francisco &52 (30-50\%)& 51 (30-60\%) & 50-51 (30-60\%) & 50 (30-50\%) \\
  Denver &52 (50-99\%)& 52 (50-80\%) & 51-52 (40-85\%) & 51 (40-60\%)  \\
  Miami &50 (55-80\%& 50-51 (40-80\%) & 50 (10-99\%) & 50 (5-65\%) \\
  Atlanta &49-50 (80-99\%)&49 (80-99\%) & 49 (90-99\%) & 49 (25-95\%) \\
  Chicago &50 (45-95\%)&49 (55-95\%) & 49 (55-95\%) & 49 (25-65\%) \\
  St. Louis &50 (70-99\%)& 51 (80-95\%)  & 50 (80-95\%) & 50 (35-95\%) \\
  New York City &52 (30-60\%)&51-52 (30-60\%) & 51 (30-60\%) & 52-53 (25-60\%) \\
  Memphis &50 (70-99\%)&50 (45-99\%) & 50 (10-90\%) & 49-50 (15-55\%) \\
  Dallas &50 (80-95\%)&50 (65-90\%) & 49 (15-85\%) & 49 (40-80\%) \\
  Houston &50 (70-90\%)&50 (80-95\%) & 50 (60-70\%) & 49-50 (30-70\%) \\
  Seattle &52 (20-65\%)&51-52 (20-50\%) & 51-52 (20-45\%) & 51 (0-50\%)\\
 \hline
 \end{tabular}
 \end{center}
\end{table}

\bigskip
\textbf{Acknowledgments}

\bigskip

\myacknow

{}
{\clearpage}
\bibliographystyle{apalike} 
\bibliography{week48bib}

\end{document}